\documentclass[11pt]{article}
\usepackage{amsmath,amsfonts}
\usepackage{tikz}
\usetikzlibrary{trees,er,snakes,shapes,mindmap}
\def\sloppy{\tolerance=100000\hfuzz=\maxdimen\vfuzz=\maxdimen}
\def \beq  {\begin{equation}}
\def \eeq  {\end{equation}}
\def \beqar {\begin{eqnarray}}
\def \eeqar {\end{eqnarray}}
\def\sqr#1#2{{\vcenter{\vbox{\hrule height.#2pt
\hbox{\vrule width.#2pt height#1pt \kern#1pt
\vrule width.#2pt}\hrule height.#2pt}}}}

\def\Tr {{\rm Tr}}

\def\del {\partial}

\def\e {\epsilon}

\def\l {\lambda}

\begin{document}
%%%%%%%%%%%%%%%%%%%%%%%%%%%%%%%%%%%%%%%%%%%%%%%
%\fontfamily{pnb}\fontsize{12pt}{16pt}\selectfont
%\fontfamily{pzc}\fontsize{14pt}{16pt}\selectfont
%\fontfamily{pbk}\fontsize{12pt}{16pt}\selectfont
\fontfamily{cmr}\fontsize{11pt}{17.2pt}\selectfont
%\fontfamily{phv}\fontshape{ro}\fontsize{11pt}{14pt}\selectfont
%\fontfamily{ptm}\fontseries{m}\fontshape{r}\fontsize{12pt}{16pt}\selectfont
%\fontfamily{pnc}\fontseries{m}\fontshape{r}\fontsize{11pt}{15pt}\selectfont
%\fontfamily{ppl}\fontseries{m}\fontshape{r}\fontsize{11pt}{15pt}\selectfont
%\usefont{T1}{phv}{m}{it}
%%%%%%%%%%%%%%%%%%%%%%%%%%%%%%%%%%%%%%%%%%%%%%%
\def \CMP {{Commun. Math. Phys.}}
\def \PRL {{Phys. Rev. Lett.}}
\def \PL {{Phys. Lett.}}
\def \NPBProc {{Nucl. Phys. B (Proc. Suppl.)}}
\def \NP {{Nucl. Phys.}}
\def \RMP {{Rev. Mod. Phys.}}
\def \JGP {{J. Geom. Phys.}}
\def \CQG {{Class. Quant. Grav.}}
\def \MPL {{Mod. Phys. Lett.}}
\def \IJMP {{ Int. J. Mod. Phys.}}
\def \JHEP {{JHEP}}
\def \PR {{Phys. Rev.}}
\def \JMP {{J. Math. Phys.}}
\def \GRG{{Gen. Rel. Grav.}}
%%%%%%%%%%%%%%%%%%%%%%%%%%%%%%%%%%%%%%%%%%%%%%%
%%%%%%%%%%%%%%%%%%%%%%%%%%%%%%%%%%%%%%%%%%%%%%%
\begin{titlepage}
\null\vspace{-62pt} \pagestyle{empty}
\begin{center}
\rightline{CCNY-HEP-10/3}
\rightline{May 2010}
\vspace{1truein} {\Large\bfseries
A Note on Schwinger Mechanism and a Nonabelian}\\
\vspace{6pt}
{\Large \bfseries Instability in a Nonabelian Plasma}\\
\vskip .1in
{\Large \bfseries  ~}\\
\vskip .1in
{\Large\bfseries ~}\\
%%%%%%%%%%%%%%%%%%%%%%%%%%%%%%%%%%%%%%%%%%%%%%%%\vspace{.6in}
{\large V.P. NAIR and ALEXANDR YELNIKOV}\\
\vskip .2in
{\itshape Physics Department\\
City College of the CUNY\\
New York, NY 10031}\\
\vskip .1in
\begin{tabular}{r l}
E-mail:
&{\fontfamily{cmtt}\fontsize{11pt}{15pt}\selectfont vpn@sci.ccny.cuny.edu}\\
&{\fontfamily{cmtt}\fontsize{11pt}{15pt}\selectfont oyelnykov@ccny.cuny.edu}
\end{tabular}

\fontfamily{cmr}\fontsize{11pt}{15pt}\selectfont
\vspace{.8in}
%\vspace{1.5in}%\vspace{0.3in}
%%%%%%%%%%%%%%%%%%%%%%%%%%%%%%%%%%%%%%%%%%%%%%%%%%%%%%%%%%%%
\centerline{\large\bf Abstract}
\end{center}
We point out that there is a nonabelian instability for a nonabelian plasma which does not 
allow both for a net nonzero color charge and the existence of field configurations which are coherent over a volume $v$ whose size is determined by the chemical potential.
The basic process which leads to this result is the Schwinger decay of chromoelectric fields, for the case where the field arises from commutators of constant potentials, rather than
as the curl of spacetime dependent potentials. In terms of the fields, instability is obtained
when $\Tr ( D^\alpha F^{\mu\nu}  \, D_\alpha F_{\mu\nu} ) > 0$.
\end{titlepage}

%%%%%%%%%%%%%%%%%%%%%%%%%%%%%%%%%%%%%%%%%%%%%%%%%%%%%%
\pagestyle{plain} \setcounter{page}{2}

The identification of the deconfined phase of quarks and gluons at the Relativistic Heavy Ion Collider, a phase akin to a nonabelian plasma, has led to a number of investigations
on instabilities in a nonabelian plasma \cite{all, all2}.
While some of these are concerned about an upgraded version of instabilities in an abelian plasma, such as the Weibel instability, there have been numerical studies of
the evolution of instabilities in the hard thermal loop approximation and beyond.
The purpose of this note is to point out that there is an instability, and a certain no-go statement, which is quite general and
arises purely from nonabelian effects.
It is fairly straightforward to understand how this effect arises.
For a statistical distribution of nonzero color charge, we need a chemical potential.
Because the charge is nonabelian in nature, the chemical potential is a matrix in the Lie algebra of the color group. In fact, it may be viewed as a background value for the time-component of the potential $A_0 = -i t^a A^a_0$, where $t^a$ form an orthonormal basis for the Lie algebra
of the color group $G$.
(We may actually take this matrix to be diagonal, but it is not important at this stage.)
If we have a constant background $A_0$, then there is an electric field generated via the commutator term $[A_0, A_i]$ in the field strength tensor.
For modes of $A_i$ of wavelength $\lambda$, this gives an electric field approximately constant over this length scale.
This electric field will then develop a Schwinger instability decaying via pair production. If the particles which are produced have a mass, there is an exponential suppression, but in the nonabelian plasma, we have effectively massless modes.
The end result of this argument is the following.
Consider the plasma coarse-grained over a distance scale $\lambda$.
Then one possibility is that the color charge density is zero when coarse-grained over this scale.
The other possibility is that the plasma cannot have $A_i$ which are coherent over length scales exceeding $\lambda$.
This is the essence of our no-go statement.

The possibility of color charge density being zero has been studied in the context of color superconductivity \cite{alford}. In the limit of large baryon number density, we expect a color superconducting phase and it is important to have color neutrality. Such a requirement can be imposed on analyses of color superconductivity, but
how it is achieved is really a dynamical issue. (This is not the setting for our question. We are concerned about a deconfined state, not superconducting and for us the baryon chemical potential can be zero. But there are points of connection.)
Nonzero charges can lead to large electric fields which are unstable, can lead to energy being nonextensive and this is one reason why stable matter must be neutral under gauge charges \cite{alford}.
Nevertheless, it is interesting to analyze some of the nuances of how neutrality is achieved.
Since the chemical potential may be taken as a background value for $A_0$, the corresponding equation of motion (or integration of the constant mode of $A_0$ in the functional integral) seems to imply zero color charge.
Strictly speaking this argument needs to be qualified, 
since it is equivalent to imposing the Gauss law
integrated over functions which do not vanish at spatial infinity.
The true gauge transformations of the theory go to the identity element at spatial infinity
and so test functions for the Gauss law must vanish at infinity.
Imposing the Gauss law with constant values for the gauge parameters is equivalent to
eliminating all charged states by fiat, which we do not want to do.
One can use a compact spatial manifold and then approach the limit of large volumes
to preserve the zero charge condition. This provides a method for carrying out the analyses,
including many of the calculations in the literature,
but it is not quite an explanation.
All this makes it useful to ask the question we are asking: If we have a deconfined state of gluons (and may be quarks), and we try to have nonzero color charge, what instabilities can arise?

The density matrix for a statistical distribution in equilibrium is given by
$\rho = \exp [ - (H- \sum_i \mu_i \, Q_i)/T]$
where $H$ is the Hamiltonian, $Q_i$ are conserved charges, $\mu_i$ are the corresponding chemical potentials and $T$ is the temperature. We are interested in time-dependent processes in this distribution, so we are concerned with real-time propagators and vertices averaged over states with the density matrix $\rho$. The result is equivalent to calculations at zero chemical potential, but with a Hamiltonian $H -\sum_i \mu_i\, Q_i$. Since the constant mode of $A_0$ couples to
$Q$,it is clear that we can treat $\mu$ as a background value for $A_0$.
Consider now the nonabelian charge density due to quarks, say, $J^a_0 = {\bar q} \gamma^0 t^a q$, or its matrix version, $(J_0)_{ij} = {\bar q}_i \gamma^0 q_j$, $i, j$ being color labels for the quarks. Under  a gauge transformation $g (x) \in G$, this matrix changes as 
\beq
J_0 \rightarrow J^g_0 =  g^{-1} ~J_0~ g
\label{1}
\eeq
It is thus possible to choose $g(x)$ such that $J_0$ is diagonal at each point.
In other words, the gauge-invariant information contained in $J_0$ may be taken as the diagonal charge densities. Thus, to specify a charge distribution, we need only chemical potentials for the Cartan elements of the Lie algebra. There are other ways to see this as well. For example, if the charged particles form some irreducible representation $R$ (which may be thought of as arising from the decomposition of a product of the representations of the individual particles), then we know that such a representation can be obtained by quantizing the co-adjoint orbit action
\beq
S = i \int dt \sum_k w_k \Tr ( h_k g^{-1}{\dot g} )
\label{2}
\eeq
where $w_k$ are the highest weights defining the representation $R$ and $h_k$ are the diagonal generators of the Lie algebra. We see that the diagonal charges are sufficient for our purpose.
In a statistical distribution, we have to think of such a representation for the global color charge over each coarse-grained volume element, and this action can be generalized to obtain the fluid flow equations for color charge \cite{bistrovic}.

In the case of a nonabelian plasma, there is an added complication. While it is possible to define a gauge-covariant charge density for the quarks (and other matter particles), there is no gauge-covariant charge density for the gluons. The integrated total charge has a gauge-invariant 
expression. The chemical potential, introduced as a background value for $A_0$ does couple to this global charge correctly. This also leads to terms quadratic in $\mu$ in the action, which is to be expected since the current for a charged bosonic system depends on $A_\mu$ in addition to the charged fields themselves. All these effects are included in the replacement $A_0 \rightarrow A_0 + \mu$. Since the diagonalization of the charge density happens only by choice of
$g(x)$, the general ansatz for the background value of $A_0$ is
\beq
A_0 = g^{-1} \mu ~g + g^{-1}\del_0 g 
\label{3}
\eeq
The group element $g$ can be removed by an overall gauge transformation,
\beqar
A_0 &\rightarrow &g A_0 g^{-1} - \del_0 g g^{-1} = \mu\nonumber\\
A_i &\rightarrow& g A_i g^{-1} - \del_i g g^{-1}
\label{4}
\eeqar
Designating the new spatial components of the potential as $A_i$ again, we see that we can
use $\mu$ as the background value for $A_0$.

\noindent $\underline{Calculating ~the ~effective~ Lagrangian}$

We shall carry out the calculations in Euclidean space. While this is not necessary, as for many other calculations at finite temperature and density, this is slightly simpler. This means that
the background value of the $A_0$ becomes imaginary.
Thus the basic calculation to check for instability reduces to the following.
Taking constant matrices for $A_0$ and $A_i$ as the background values, we consider fluctuations in the fields. 
The integration of the action to quadratic order in the fluctuations leads to the standard
determinant.
This has to be evaluated as a function of the background values.
The result is then analytically continued to imaginary values of the background $A_0$.
The result can then be analyzed for instbilities.
The instability of interest to us is the Schwinger decay of the chromoelectric field.
This has been studied in some detail in the nonabelian case for electric fields which are given by the curl of the gauge potentials \cite{nayak}, but, here, we are interested in the case when the field arises from the commutator term of the potentials.
For the calculations which follow,
we will consider the group $SU(2)$ since it is sufficient to capture the effect we are interested in.

The integration over the quadratic fluctuations can be phrased as an effective Lagrangian given by

\beqar
L_{eff} &=& {1\over 2} \int {d^Dp\over (2\pi )^D} \int_0^\infty {ds \over s}
\Tr \left[ \exp \left( -s [-(D^2)^{ab}\eta_{\mu\nu} - 2 f^{acb} F_{\mu\nu}^c]\right)\right]\nonumber\\
&& \hskip .2in - \int {d^Dp\over (2\pi )^D} \int_0^\infty {ds \over s}
\Tr \left[ \exp \left( -s [-D^2 ]\right)\right]\label{5}
\eeqar
where the second term is the contribution from the ghosts.
Here $D^2= (\del_\mu +A_\mu ) (\del^\mu + A^\mu)$ is the gauge-covariant Laplacian with the background field $A^a_\mu$; it is a $3\times 3$-matrix in color space, as indicated by the
color indices
$a, b$.
Thus the operator $-(D^2)^{ab} \eta_{\mu\nu} - 2 f^{acb} F_{\mu\nu}^c$ can be considered as a 
$12\times 12$-matrix, in addition to its coordinate space properties.
The evaluation of the action will follow a method which is similar to what was used many years ago by Brown and Weisberger \cite{BW}.
Writing the $SU(2)$ field $A^{ab}_\mu = f^{acb}A^c_\mu = \epsilon^{acb} A^c_\mu$, we can simplify $D^2$ as
\beq
- (D^2)_{ab} = p^2 + Y - Y_{ab} - 2 i p\cdot A_{ab}
\label{6}
\eeq
where $p_\mu = - i \del_\mu$,  $Y^{ab} = A^a_\mu A^b_\mu$ and $Y = \Tr Y^{ab}$. 
The matrix $Y^{ab}$ can be diagonalized by a suitable gauge transformation, with eigenvalues
$\lambda_a$.  These eigenvalues give the gauge-invariant characterization of the chromoelectric
and chromomagnetic fields.
The $\lambda$'s are positive in the case of Euclidean signature for the
contraction of spacetime indices in $ A^a_\mu A^b_\mu$, but one eigenvalue can be negative
with Minkowski signature. 
In the Euclidean metric we are using, we can always choose 
\beqar
A^a_\mu &=& \sqrt{\lambda_a }~\delta^a_\mu, \hskip .2in a, \mu = 1,2, 3\nonumber\\
A^a_4&=&  0
\label{7}
\eeqar
With this choice
\beq
Y_{ab} + 2 i (p\cdot A)_{ab} = \left[
\begin{matrix}
\lambda_1 & -2i p_3 \sqrt{\lambda_3}& 2i p_2 \sqrt{\lambda_2}\\
2i p_3 {\sqrt{\lambda_3}}& \lambda_2& - 2i p_1 \sqrt{\lambda_1}\\
- 2i p_2 \sqrt{\lambda_2}& 2i p_1 \sqrt{\lambda_1}&\lambda_3\\
\end{matrix}
\right]
\label{8}
\eeq
For our purpose, it is not necessary to consider this matrix in full generality, we can take 
$\lambda_3 =0$. In this case the only nontrivial component of the field strength tensor is
$F_{12}^3 = -F_{21}^3 = \sqrt{\lambda_1 \lambda_2}$. In this case, schematically, we have
\beqar
\left[ Y_{ab} + 2 i (p\cdot A)_{ab}\right] \eta_{\mu\nu}  + 2 F_{ab\mu\nu}
= \left[ 
\begin{matrix}
Y+ 2ip\cdot A &2F_{12} &0&0\\
-2 F_{12}&Y+2ip\cdot A&0&0\\
0&0&Y+2 i p\cdot A& 0\\
0&0&0&Y+2 i p\cdot A\\
\end{matrix}
\right]
\label{9}
\eeqar
where each block is a $3\times 3$ matrix in color space.
From this block diagonal form, 
\beqar
\Tr_{12\times 12} \exp \left[ s \left\{ ( Y+2i p\cdot A )\eta_{\mu\nu} + 2 F_{\mu\nu}\right\} \right]
&=& 2 ~\Tr_{3\times 3} ~e^{s(Y+ 2ip\cdot A)} \nonumber\\
&&\hskip .2in +~
\Tr_{6\times 6} ~e^{s[ (Y+2ip\cdot A )\eta_{\mu\nu} + 2 F_{\mu\nu} ]}
\label{10}
\eeqar
The first term on the right hand side cancels exactly the similar contribution from 
ghosts. The remaining $6\times 6$ matrix corresponds to the indices $1, 2$, for spacetime and the 
$3\times 3$ matrix in color space. The effective Lagrangian is thus
\beq
L_{eff} = {1\over 2} \int {d^Dp\over (2\pi )^D} \int_0^\infty {ds \over s}
e^{-s (p^2 +Y)}~ \Tr_{6\times 6} e^{-s {\mathbb X}}
\label{11}
\eeq
where ${\mathbb X}$ is the $6\times 6$ matrix
%\beq
%(- {\mathbb X}) =
%\left[ 
%\begin{matrix}
%Y+ 2ip\cdot A &2F_{12} \\
%-2 F_{12}&Y+2ip\cdot A\\
%\end{matrix}
%\right]
%\label{11a}
%\eeq
\beq
(- {\mathbb X}) =
 \left[ 
\begin{matrix}
\lambda_1 & 0 & 2 i p^2 \sqrt{\lambda_2} & 0 & -2\sqrt{\lambda_1\lambda_2} & 0 \\
0 &\lambda_2 & -2 i p^1 \sqrt{\lambda_1} & 2\sqrt{\lambda_1\lambda_2}& 0 & 0\\
-2 i p^2 \sqrt{\lambda_2} &  2 i p^1 \sqrt{\lambda_1} & 0 & 0 & 0 & 0\\
0 &  2\sqrt{\lambda_1\lambda_2} & 0 & \lambda_1 & 0 & 2 i p^2 \sqrt{\lambda_2}\\
-2\sqrt{\lambda_1\lambda_2} & 0 & 0 & 0 & \lambda_2 & -2 i p^1 \sqrt{\lambda_1}\\
 0 & 0 & 0 & -2 i p^2 \sqrt{\lambda_2} &  2 i p^1 \sqrt{\lambda_1} & 0\\
\end{matrix}
\right]
\label{11a}
\eeq
For evaluating the remaining trace, it is convenient to use the integral representation
\beq
\Tr e^{-s{\mathbb X}}
= \oint {dz \over 2\pi i} e^{-sz} {\del \over \del z} \log \det (z- {\mathbb X})
\label{12}
\eeq
where the integration contour encircles all zeros of $\det (z - {\mathbb X})$.

The determinant is easy to evaluate,
\begin{subequations}
\beqar
\hskip -.15in \det (z - {\mathbb X}) &\!\!\!=\!\!\!&\left\{ z^3 + z^2 (\lambda_1 + \lambda_2) -  \left[ 4 p_1^2 (z \lambda_1 +\lambda_1^2)
+4 p_2^2 (z \lambda_2 + \lambda_2^2) + 3 z \lambda_1 \lambda_2\right]\right\}^2\label{13a}\\
&\!\!\!=\!\!\!& \left\{z \left[ z^2 + z (\l_1 +\l_2) - 3\l_1 \l_2\right]
\left[ 1 - {4p_1^2 ( z \l_1+ \l_1^2) +4p_2^2 ( z \l_2+ \l_2^2)
\over z \left[ z^2 + z (\l_1 +\l_2) - 3\l_1 \l_2\right] }\right]\right\}^2~~
\label{13b}
\eeqar
\end{subequations}
When this is used in (\ref{11},\ref{12}), with the $\del_z$ carried out, we get contributions from the poles which correspond to the roots of the cubic polynomial inside the braces in (\ref{13a}). It is then convenient to split the expression for $L_{eff}$ as $L_1 +L_2$ with
\begin{subequations}
\beqar
L_1 &=&  \int {d^Dp\over (2\pi )^D} \int_0^\infty {ds \over s} e^{-s(p^2 +\l_1 +\l_2)}\nonumber\\
&&\hskip .3in \times \oint {dz \over 2\pi i} e^{-sz} {\del \over \del z} \log \left[
z \left(z^2 +z (\l_1+\l_2) - 3 \l_1 \l_2\right)\right]
\label{14a}\\
L_2&=& \int {d^Dp\over (2\pi )^D} \int_0^\infty {ds \over s} e^{-s(p^2 +\l_1 +\l_2)}\nonumber\\
&&\hskip .3in \times \oint {dz \over 2\pi i} e^{-sz} {\del \over \del z} \log \left[
1 - {4p_1^2 ( z \l_1+ \l_1^2) +4p_2^2 ( z \l_2+ \l_2^2)
\over z \left[ z^2 + z (\l_1 +\l_2) - 3\l_1 \l_2\right] }\right]
\label{14b}
\eeqar
\end{subequations}

The evaluation of $L_1$ is simple. The zeros of the relevant cubic polynomial are $z=0$ and
$z= z_\pm$ with
\beq
z_\pm = {1\over 2} \left[ -(\l_1 + \l_2) \pm \sqrt{(\l_1 +\l_2)^2 + 12 \l_1 \l_2}\right]
\label{15}
\eeq
We then find
\beq
L_1 =  {\Gamma (-D/2) \over (4\pi )^{D/2}}
\biggl[ (\l_1 + \l_2)^{D/2} + \left( \l_1 + \l_2 + z_+ \right)^{D/2} + \left( \l_1 + \l_2 + z_-\right)^{D/2} \biggr]
\label{16}
\eeq
$\Gamma$ is the Eulerian gamma function. Notice that there are singularities in this
expression for $D=4$. These are, of course, the standard renormalization singularities
and can be isolated by expanding $(\mu )^{4 -D} L_1$ in powers of $\e$ with $D = 4- \e$.
(The $\mu$-factor is the usual one for ensuring the correct dimension for $L_1$.)
This leads to the expression
\beqar
\mu^{4-D} L_1 &=& {1\over (4\pi)^2 \e} \left[ (\l_1 +\l_2)^2 + (\l_1 +\l_2 +z_+)^2
+ (\l_1 +\l_2 +z_-)^2 \right]\nonumber\\
&&+{(\l_1 +\l_2)^2\over (4\pi )^2} \left({3\over 4} - {1\over 2} \log (\l_1 +\l_2)/{\tilde \mu}^2\right)
\nonumber\\
&&+{(\l_1 +\l_2+z_+)^2\over (4\pi )^2} \left({3\over 4} - {1\over 2} \log (\l_1 +\l_2 +z_+)/{\tilde \mu}^2\right)\nonumber\\
&&+{(\l_1 +\l_2 +z_-)^2\over (4\pi )^2} \left({3\over 4} - {1\over 2} \log (\l_1 +\l_2 +z_-)/{\tilde \mu}^2\right) ~+~{\cal O}(\e ) \label{26}
\eeqar
where ${\tilde \mu }^2= 4 \pi e^{-\gamma} \mu^2$, $\gamma$ being the Euler-Mascheroni constant.

The first term on the right hand side of (\ref{26}) is the potentially divergent contribution which is removed
by renormalization. The remainder gives the finite expression we need for $L_1$.

The evaluation of $L_2$ is a little more involved and is sketched out in the appendix.
The final result is
\beq
L_2 = - {1\over (4\pi )^{D/2} \Gamma (D/2)} \int_0^\infty dz \int_0^1 dx
{(1-x)^{-1+D/2} \over x} z^{-1+D/2}  \left[ 1 - {1\over \sqrt{1- x A_1}\, \sqrt{1- x A_2}}
\right]\label{27}
\eeq
where
\beq
A_1 = { 4 z \l_1 (z +\l_2) \over (z+\l_1 +\l_2) \left[ (z+\l_1) (z+\l_2) - 4 \l_1 \l_2\right]}
\label{27a}
\eeq
and $A_2$ is obtained by the exchange $\l_1 \leftrightarrow \l_2$ in the above expression.
In (\ref{27}) also,  there is a potentially divergent contribution arising from the large $z$ behavior of the integrand. Its removal, along with the potentially divergent term from
(\ref{26}) is discussed in the appendix.

\noindent $\underline{The ~nature ~of ~the ~instability}$

We are now in a position to consider how instabilities can arise from these results.
In continuing the expressions for $L_1$, $L_2$ to Minkowski space, one of the directions has to be identified as the time-direction. We will take this to be the $1$-direction. Chromoelectric fields in Minkowski space will thus correspond to the choice $\l_1 <0$, $\l_2 >0$. The choice of
$\l_1, \l_2 >0$ will correspond to the purely chromomagnetic case, with $1$-direction being interpreted as spatial direction now.
We will consider various possibilities for the $\lambda$'s one by one.

\noindent{$\underline{Case~a}$}: \\
Consider first the case of $\l_1 <0$, $\l_2 >0$, $\l_1 + \l_2 >0$. In this case,
the factor $(\l_1 + \l_2 )^2 + 12 \l_1 \l_2$ is positive for $\l_2 \gg \vert\l_1 \vert$. For this region
\beq
\l_1 + \l_2 + z_\pm = {\l_1 + \l_2 \pm \sqrt{(\l_1 +\l_2)^2 + 12 \l_1 \l_2 } \over 2} ~> 0
\label{28}
\eeq
and hence there is no instability in $L_1$. 
As we come down in the value of $\l_2$, this factor changes sign at $\l_2 = (7 +\sqrt{48}) \vert \l_1\vert$. For the region $\vert \l_1\vert < \l_2 < (7 +\sqrt{48}) \vert \l_1 \vert$, the quantities
$\l_1 +\l_2 + z_+$ and $\l_1 +\l_2 + z_-$ are complex conjugates of each other. Writing these
as $ \alpha e^{\pm i \theta}$, we can easily see from (\ref{26}) that there is no imaginary part in
$L_1$ for this region as well.
Thus, there is no instability resulting from $L_1$. 

Turning to ${\rm Im} L_2$, notice that we can set  $D =4$ at this stage because the integration range for $z$ for the imaginary part does not extend to infinity and so the issue of divergences do not arise.
The analysis of $L_2$ then reduces to the analysis of the condition $A_2 (z) > 1$.
The polynomial factor in the denominator of the $A$'s, namely, that $(z+\l_1) (z +\l_2 ) - 4 \l_1 \l_2 = z^2 + z (\l_1 +\l_2) + 3 \vert \l_1\vert \l_2 $ is easily seen to be positive.
Thus $A_1 (z) < 0$ and the factor $\sqrt{1 - x A_1}$ is real for the full range ($z>0$) of
integration for $z$.
On other hand, $A_2(z)$, whose numerator is $4 z \l_2 (z +\l_1)$ will show a change of sign for $z = -\l_1 > 0$. However, even though $A_2 (z) > 0$ for $z > \l_1$,  we have $A_2 (z) \leq 1$.
This is easily seen from the fact that 
\beq
(z+\l_1 +\l_2) \left[ (z+\l_1) (z+\l_2) - 4 \l_1 \l_2\right] 
\geq  (z- \vert \l_1\vert) \left[ (z - \vert\l_1\vert) (z +\l_2) + 4 \vert \l_1 \vert  \, \l_2
\right]\label{29}
\eeq
The quantity in the square brackets on the right hand side is $\geq 4 z \l_2$ for
$z > \vert \l_1 \vert$.
Thus the factor $\sqrt{1 - x A_2 }$ is also real and hence there is no instability for this case from either $L_1$ or $L_2$.

\noindent{$\underline{Case ~b}$}: \\
Now we turn to the case $\l_1 <0$, $\l_2 >0$, $\l_1 + \l_2 < 0$. 
The region $(7 - \sqrt{48})\vert \l_1\vert < \l_2 < \vert \l_1\vert$ has complex conjugate
values for $\l_1 + \l_2 +z_\pm$ and there is no imaginary part resulting from the last two terms in  $L_1$, as in the previous case for $\l_2 < (7 + \sqrt{48})\vert \l_1 \vert$.
There is an imaginary part from the $\log (\l_1 + \l_2)$ term in $L_1$,
which will give an instability for this range of $\l_2$.
For $\l_2 < (7 - \sqrt{48}) \vert \l_1 \vert$ (or $\vert \l_1 \vert > (7 +\sqrt{48}) \l_2$)
we have
\beq
\l_1 + \l_2 + z_\pm =  {\l_1 + \l_2 \pm \sqrt{(\l_1 +\l_2)^2 + 12 \l_1 \l_2 } \over 2} ~< 0
\label{30}
\eeq
There is then a nontrivial imaginary part in $L_1$ which leads to an instability.
Thus we get instability from $L_1$ for all $\l_1$, $\l_2$ corresponding to this case.

\begin{figure}[!t]
\begin{center}
\scalebox{.8}{\includegraphics{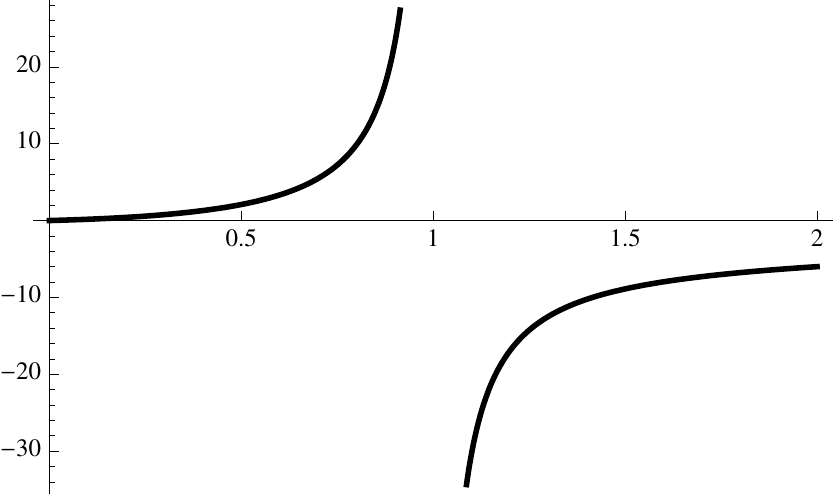}}\hskip .3in
\scalebox{.8}{\includegraphics{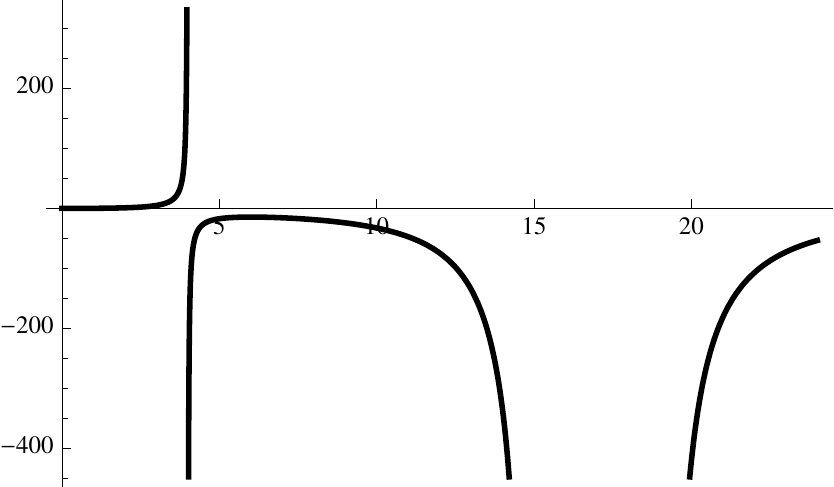}}
\vskip .35in
\caption{Sample graphs of $A_1(z)$ for $1< \vert \l_1\vert/\l_2 < 7+\sqrt{48}$ (left)
and for $\vert \l_1\vert/\l_2 > 7+\sqrt{48}$ (right). The value of $A_1$ between
$15$ and $20$ is large and positive and outside the frame of the graph on the right.}\label{fig1}
\end{center}
\end{figure}

\begin{figure}[!t]
\begin{center}
\scalebox{.8}{\includegraphics{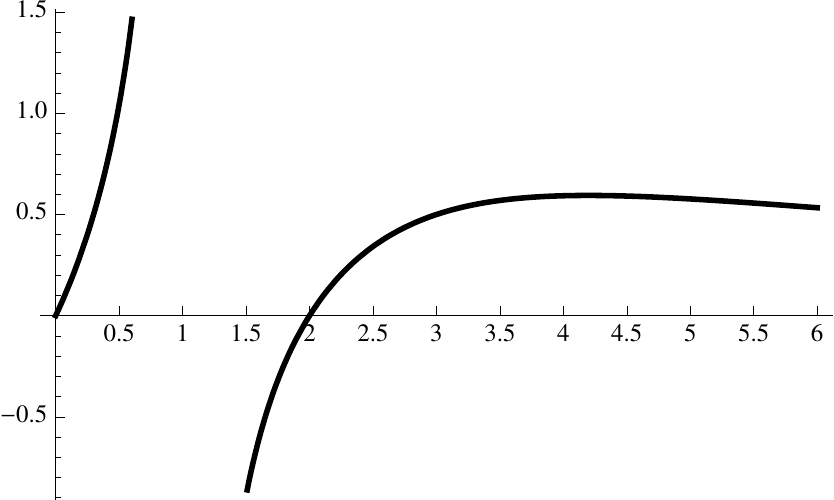}}\hskip .3in
\scalebox{.8}{\includegraphics{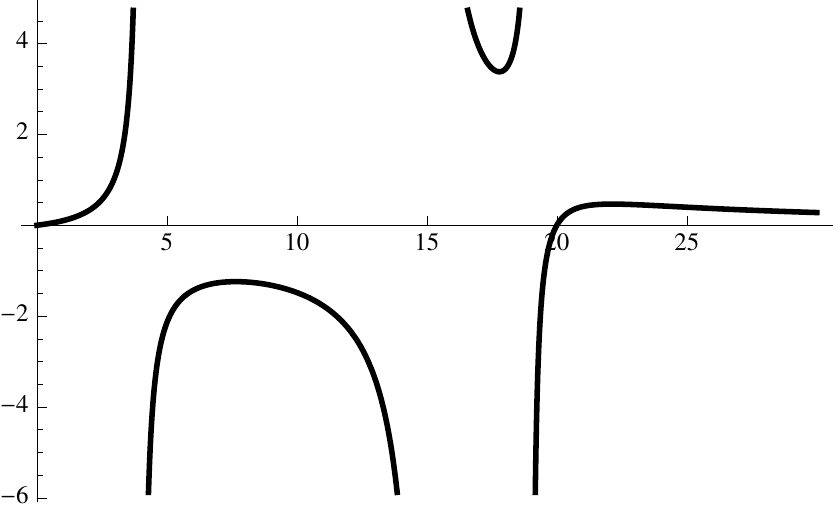}}
\vskip .35in
\caption{Sample graphs of $A_2(z)$ for $1< \vert \l_1\vert/\l_2 < 7+\sqrt{48}$ (left)
and for $\vert \l_1\vert/\l_2 > 7+\sqrt{48}$ (right). }\label{fig2}
\end{center}
\end{figure}

Turning to $L_2$, we may notice that the factor  $(z +\l_1 +\l_2)$
 in the denominator of $A_1$, $A_2$ changes sign
at $z = - (\l_1 +\l_2 )$.  The additional factor in the denominator, namely,
$ [(z+\l_1) (z+\l_2) - 4 \l_1 \l_2]$ has two positive roots if $\vert \l_1\vert /\l_2 >
7 +\sqrt{48} \approx 14$. Otherwise, there are no real roots and this factor is positive.
The graphs of $A_1(z)$ as  a function of $z$ are as shown in Fig.\ref{fig1}.
We see that for all values of $\vert \l_1 \vert /\l_2$, there are regions of $z$-integration for which $A_1(z) > 1$, leading to
an imaginary part for $L_2$. Similar statements apply for $A_2$, see Fig.\ref{fig2}.
%%%%%%%%%%%%%%%%%%%%%%%%%%
%%%%%%%%%%%%%%%%%%%%%%%%%%%%
%%%%%%%%%%%%%%%%%%%%%%%%%%
%%%%%%%%%%%%%%%%%%%%%%%%%%%%

\pagebreak
\noindent{$\underline{Case ~c}$}:\\
Even though it is not germane to our present discussion, we may note that if we have the purely chromomagnetic case with $\l_1 > 0$, $\l_2 > 0$, then
\beq
\l_1 + \l_2 + z_-  = {(\l_1 +\l_2) - \sqrt{(\l_1 + \l_2 )^2 + 12 \l_1 \l_2 } \over 2} < 0
\label{30a}
\eeq
Thus the last term
on the right hand side in (\ref{26}) has an imaginary component. For $L_2$, the 
polynomial $(z+\l_1) (z+\l_2) - 4 \l_1 \l_2$ in the denominators of $A_1$, $A_2$ has roots $z_\pm$.
For $\l_1 , \l_2 > 0$, one root is negative and the other is positive. 
$A_1(z) $ is positive for $z > z_+$ and goes to zero
for large $z$, with  $A_1 (z) \rightarrow \infty $ for $z - z_+ \rightarrow 0_+$.
Thus there is a range of $z$ for which $\sqrt{1 - x A_1}$ has an imaginary part.
Again a similar statement applies to $A_2$. Thus for both $L_1$ and $L_2$ we get an instability
for $\l_1, \l_2 > 0$.
This is the well-known vacuum instability in a chromomagnetic field.

It is interesting to characterize the instability in terms of invariants of the field.
We see easily that $F^a_{\mu\nu} F^{a \mu\nu} = (\Tr Y)^2 - \Tr Y^2 = 2 \l_1 \l_2$,
$(D^\alpha F^{\mu\nu})^a (D_\alpha F_{\mu\nu})^a = 2 \l_1 \l_2 ( \l_1 +\l_2 )$.
We may then summarize our results as
\beq
\Tr ( D^\alpha F^{\mu\nu}  \, D_\alpha F_{\mu\nu} )~
\left\{~
\begin{matrix}
>& 0& \hskip .2in {\rm Instability}\\
< &0 & \hskip .35in  {\rm No ~ instability}\\
\end{matrix}
\right.
\label{31}
\eeq

\noindent{$\underline{Comments}$}

The instability we are discussing is quite general and hints at how statistical distributions
tend to move to color neutrality or a disordered state with no coherent fields over 
distances long compared to the dimension given by the chemical potential.
The calculation itself may be taken as the derivation of Schwinger decay of chromoelectric fields for the case when the field is generated by the commutator term, rather than the curl of the potentials.
\bigskip

This work was supported by U.S.\ National Science
Foundation grant PHY-0855515
and by a PSC-CUNY award.

\section*{\small APPENDIX}
\noindent $\underline{Calculation ~ of ~ L_2}$

\begin{figure}[!t]
\begin{minipage}{6cm}
\begin{center}
\scalebox{.8}{\includegraphics{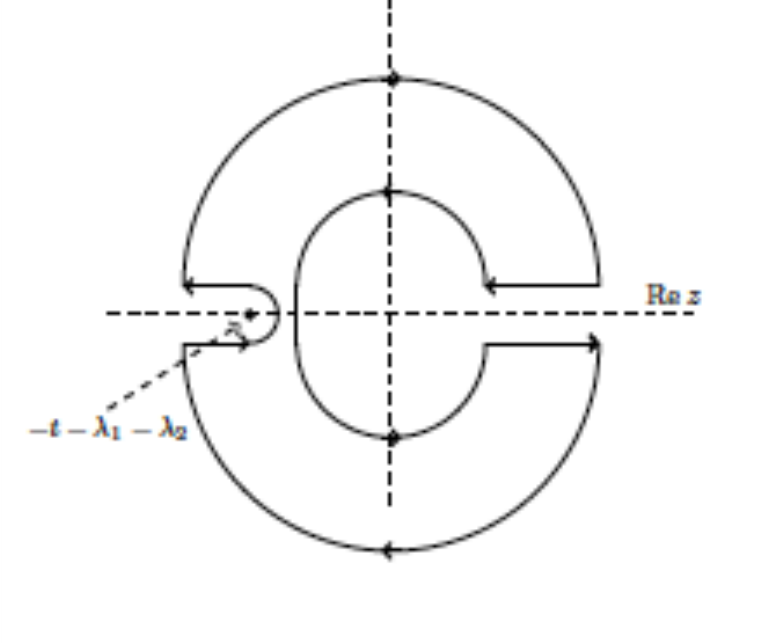}}\label{fig3}\\
\vskip .1in
\caption{Deformation of contour for branch cut at $z= -t -\l_1 -\l_2$}\label{fig3}
\end{center}
\end{minipage}
\hskip 1in
\begin{minipage}{6cm}
\begin{center}
\scalebox{.9}{\includegraphics{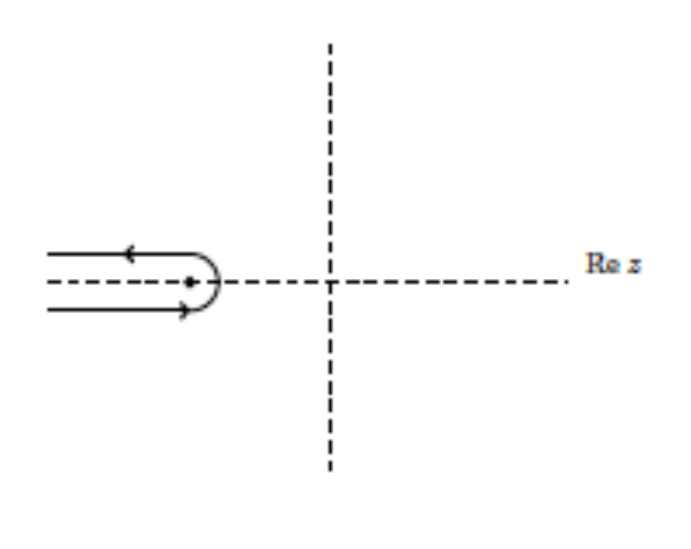}}\\
\vskip .35in
\caption{Contour for evaluating $L_2$}\label{fig4}
\end{center}
\end{minipage}
\end{figure}

For $L_2$, we start with the representation
\beq
\log A = \int_0^\infty {dt \over t} (e^{-t} - e^{-tA})
\label{17}
\eeq
Using this and eliminating $\del_z$ by partial integration, the expression (\ref{14b}) for $L_2$ becomes
\beqar
L_2 &=& \int {d^Dp\over (2\pi )^D} \int_0^\infty {ds \over s} e^{-s(p^2 +\l_1 +\l_2)}\nonumber\\
&&\hskip .3in \times \oint {dz \over 2\pi i} e^{-sz} \int_0^\infty {dt \over t} e^{-st} 
\left[ 1 - \exp\left( 4st {p_1^2 \l_1 (z+\l_1) + p_2^2 \l_2 (z+ \l_2) \over z (z^2 + z\l_1 +z\l_2 - 3\l_1 \l_2)}\right)\right]\nonumber\\
&=&\int_0^\infty {ds \over (4\pi s)^{D/2}} e^{-s(z+ t+\l_1+\l_2)}
\oint {dz \over 2\pi i} \int_0^\infty {dt\over t} 
\left[ 1 - {1\over C_1(z) C_2(z)}\right]
\label{18}
\eeqar
where
\beq
C_k(z) = \sqrt{ 1- { 4t \l_k (z+\l_k) \over z (z^2 + z\l_1 +z\l_2 - 3\l_1 \l_2)}}
\label{19}
\eeq
for $k = 1, 2$.
For the second line of equation (\ref{18}) we have carried out the $p$-integration.
Note that the exponents involving $p_k^2$ show that we need o take the $z$-contour to be large enough, $\vert z\vert >  2 \sqrt{t\l}$. Effectively, this means that we should do the $z$-interal before doing the $t$-integral.
In (\ref{18}), we can further carry out the $s$-integration to get
\beq
L_2 = {\Gamma(1-D/2)\over (4\pi )^{D/2} } \int_0^\infty {dt \over t} \oint {dz \over 2\pi i} 
(z + t + \l_1 +\l_2)^{-1+D/2} \left[ 1 - {1\over C_1(z) C_2(z)}\right]
\label{20}
\eeq
The factor $(z + t + \l_1 +\l_2)^{-1+D/2} $ shows that, for the $z$-integration, we have a branch cut along the negative real axis starting at  $z = -t -\l_1 -\l_2$. We can deform the original contour
which is a large circle around the origin, via the contour shown in Fig.\ref{fig3}, to the contour in Fig.\ref{fig4}.
Notice that  because of the arguments given earlier, the branch point $z = -t -\l_1 -\l_2$
is always outside the original contour, while the singularities of the square root factors are always inside the contour.
Integration along the cut in Fig.\ref{fig4} gives
\beqar
L_2 &=& {\Gamma(1-D/2)\over (4\pi )^{D/2} } \int_0^\infty {dt \over t} 
\int_{t+\l_1+\l_2}^\infty \!\!\!\!dz (z - t -\l_1 -\l_2)^{-1+D/2} \left[ {e^{i\pi (D/2 -1)} - e^{-i\pi (D/2 -1)} 
\over 2\pi i}\right]\nonumber\\
&&\hskip 1.5in \times \left[ 1 - {1\over C_1(-z)  C_2(-z)}\right]
\label{21}
\eeqar
Using
\beq
{e^{i\pi (D/2 -1)} - e^{-i\pi (D/2 -1)} \over 2\pi i} = - {1\over \Gamma (1-D/2) \Gamma (D/2)}
\label{22}
\eeq
and shifting the variable of integration to $z-\l_1- \l_2$, we can write (\ref{21}) as
\beq
L_2 = - {1\over (4\pi )^{D/2} \Gamma (D/2)} \int_0^\infty {dt\over t} \int_t^\infty
dz (z-t)^{-1+D/2} \left[ 1 - {1\over \sqrt{1- t A_1 /z} \sqrt{1- tA_2/z}}
\right]\label{23}
\eeq
where
\beq
A_1 = { 4 z \l_1 (z +\l_2) \over (z+\l_1 +\l_2) \left[ (z+\l_1) (z+\l_2) - 4 \l_1 \l_2\right]}
\label{24}
\eeq
and $A_2$ is given by the same expression with $\l_1 \leftrightarrow \l_2$.
Changing the order of integration and making the substitution $t = z x$, we finally get
\beq
L_2 = - {1\over (4\pi )^{D/2} \Gamma (D/2)} \int_0^\infty dz \int_0^1 dx
{(1-x)^{-1+D/2} \over x} z^{-1+D/2}  \left[ 1 - {1\over \sqrt{1- x A_1}  \sqrt{1- x A_2}}
\right]\label{25}
\eeq
This is the expression quoted in the text.

\noindent $\underline{Renormalization: ~A ~consistency ~check}$

The potentially divergent part of $L_1$ was obtained in equation (\ref{26}) as
\beq
\mu^{4-D} L_{1 div} = {1\over (4\pi)^2 \e} \left[ (\l_1 +\l_2)^2 + (\l_1 +\l_2 +z_+)^2
+ (\l_1 +\l_2 +z_-)^2 \right]
\label{32}
\eeq
Using the expressions for $z_\pm$ from (\ref{15}), this simplifies to
\beq
\mu^{4-D} L_{1 div} = {1\over (4\pi)^2 \e} \left[ 2 (\l_1^2 + \l_2^2 ) + 10 \l_1 \l_2 \right]
\label{33}
\eeq
The expression for $L_2$ can be recast as
\beq
L_2 = - {1\over (4 \pi)^{D/2} \Gamma (D/2)} \int_0^\infty  d\tau ~\tau^{-1 -D/2} ~G (\tau )
\label{34}
\eeq
where $\tau = 1/z$ and
\beq
G(\tau ) = \int_0^1 {dx \over x} (1- x)^{-1 +D/2}
\left[ 1 - {1\over \sqrt{1- x {\tilde A_1}} \sqrt{1- x {\tilde A_2}} }\right]
\label{35}
\eeq
and ${\tilde A}$'s correspond to $A$'s with $z = 1/\tau$; i.e.,
\beq
{\tilde A}_1 =  {4 \tau \l_1 (1 +\tau \l_2) \over [1+ \tau (\l_1 +\l_2)] \, [ (1+ \tau \l_1 )
(1 +\tau \l_2 ) - 4 \tau^2 \l_1 \l_2 ]}
\label{36}
\eeq
with $\l_1 \leftrightarrow \l_2$ to obtain ${\tilde A}_2$ from ${\tilde A}_1$.
The divergence now corresponds to small values of $\tau$. Carrying out a small $\tau$-expansion,
\beq
G (\tau ) = - {4\over D} \tau (\l_1 + \l_2) + {8 \tau^2 \over D (D+2)}
\left[ D \l_1 \l_2 + (D-1) (\l_1^2 + \l_2^2)\right] ~+~ {\cal O}(\tau^3)
\label{37}
\eeq
We can use this expansion in (\ref{34}) and integrate; we are interested in small $\tau$ region, so we use a cutoff $e^{-\tau}$ in the integrand. (Whether we use this or something else, such as
$e^{- a \tau}$ for some $a$ does not matter for the term of the form $\Gamma ((4-D)/2)$.)
The term proportional to $1/\e$ is then found to be
\beq
L_{2 div} = - {1\over (4\pi )^2 \e} \left[ 2 (\l_1^2 + \l_2^2) + {8\over 3} \l_1 \l_2
\right]
\label{38}
\eeq
Combining this with (\ref{33}), we find
\beq
L_{div} = {1\over (4\pi )^2 \e} {22 \over 3} \l_1 \l_2 =
{1\over (4\pi )^2 \e} {11 \over 3} F^a_{\mu\nu} F^{a\mu\nu}
\label{39}
\eeq
This is the expected and correct renormalization of the action, and is consistent with the
$\beta$-function of
\beq
\beta (g) = - {g^3 \over (4\pi)^2} {22\over 3}
\label{40}
\eeq
for $SU(2)$.
%%%%%%%%%%%%%%%%
%%%%%%%%%%%%%%%%


\begin{thebibliography}{99}

\bibitem{all}  S.~Mrowczynski,
 %  {\em  Plasma instability at the initial stage of ultrarelativistic heavy ion
%  collisions,} 
  Phys.\ Lett.\ B {\bf 314}, 118(1993);
  P.~Arnold, J.~Lenaghan and G.~D.~Moore,
%{\em  QCD plasma instabilities and bottom-up thermalization,} 
JHEP {\bf 0308} (2003) 002
 [arXiv:hep-ph/0307325]; 
P.~Arnold and J.~Lenaghan,
 % {\em  The abelianization of QCD plasma instabilities, } 
  Phys.\ Rev.\ D {\bf 70} (2004) 114007
  [arXiv:hep-ph/0408052];
  P. Arnold and J. Lenaghan, \PR~{\bf D70}, 114007 (2004);
  D.~B\"odeker,
 % {\em  The impact of QCD plasma instabilities on bottom-up
 % thermalization,} 
  JHEP {\bf 0510} (2005) 092
  [arXiv:hep-ph/0508223];
    P. Arnold, J. Lenaghan, G.D. Moore and L.G. Yaffe,
%{\em Apparent thermalization due to plasma instabilities in
%quark gluon  plasma}, 
\PRL~{\bf 94}, 072302 (2005);
P.~Arnold, G.~D.~Moore and L.~G.~Yaffe,
  %{\em  The fate of non-abelian plasma instabilities in 3+1 dimensions,} 
  Phys.\ Rev.\ D {\bf 72} (2005) 054003
  [arXiv:hep-ph/0505212];
A. Dumitru and Y. Nara,\PL~{\bf B621}, 89 (2005);
B. Schenke, M. Strickland, C. Greiner and M.H. Thoma,
\PR~{\bf D73}, 125004 (2006); Y. Nara, \NP~{A774}, 783 (2006);
P. Arnold and G.D. Moore, \PR~{\bf D73}, 025006 (2006);

\bibitem{all2}
  A.~Rebhan, P.~Romatschke and M.~Strickland,
 % {\em  Hard-loop dynamics of non-Abelian plasma instabilities,} 
  Phys.\ Rev.\ Lett.\  {\bf 94} (2005) 102303
  [arXiv:hep-ph/0412016];
A.~Rebhan, P.~Romatschke and M.~Strickland,
%{\em  Dynamics of quark-gluon plasma instabilities in discretized hard-loop 
%approximation,} 
  JHEP {\bf 0509} (2005) 041
  [arXiv:hep-ph/0505261];
  D.~Bodeker and K.~Rummukainen,
 % {\em Non-abelian plasma instabilities for strong anisotropy,}
  JHEP {\bf 0707} (2007) 022
  [arXiv:0705.0180 [hep-ph]];
  P.~Arnold and G.~D.~Moore,
 % {\em Non-Abelian Plasma Instabilities for Extreme Anisotropy,}
  Phys.\ Rev.\  D {\bf 76} (2007) 045009
  [arXiv:0706.0490 [hep-ph]];
  A.~Dumitru, Y.~Nara and M.~Strickland,
%  {\em Ultraviolet avalanche in anisotropic non-Abelian plasmas,}
  Phys.\ Rev.\  D {\bf 75} (2007) 025016
  [arXiv:hep-ph/0604149].
                
  
\bibitem{alford} See, for example, M.G. Alford, K. Rajagopal, T. Schaefer and A. Schmitt, \RMP~{\bf 80}, 1455 (2008)
[arXiv:0709.4635[hep-ph]]

\bibitem{bistrovic} B. Bistrovic,
R. Jackiw, H. Li, V.P. Nair and S-Y. Pi, Phys. Rev. {\bf D67}, 025013 (2003);
 R. Jackiw, V.P. Nair, S-Y. Pi and
A.P. Polychronakos, J. Phys. A: Math. Gen. {\bf 37}, R327 (2004).

\bibitem{nayak} G. Nayak and P. van Nieuwenhuizen, \PR~{\bf D71}, 125001 (2005);
F. Cooper and G. Nayak, \PR~{\bf D73}, 065005 (2006);
G. Nayak, Eur. Phys. J. {\bf C59}, 715 (2009); 
S.P. Gavrilov and D.M. Gitman, Eur. Phys. J. {\bf C64}, 81 (2009);
G. Nayak, \IJMP~{\bf A25}, 1155 (2010).

\bibitem{BW} L.S. Brown and W.I. Weisberger, \NP~{\bf B157}, 285 (1979); \NP~{\bf B161}, 61 (1979).

\end{thebibliography}
\end{document}